# Design and implementation of self-adaptable parallel algorithms for scientific computing on highly heterogeneous HPC platforms

Alexey Lastovetsky[*], Ravi Reddy, Vladimir Rychkov, David Clarke

*School of Computer Science and Informatics, University College Dublin, Belfield, Dublin 4, Ireland*

**Abstract**

Traditional heterogeneous parallel algorithms, designed for heterogeneous clusters of workstations, are based on the assumption that the absolute speed of the processors does not depend on the size of the computational task. This assumption proved inaccurate for modern and perspective highly heterogeneous HPC platforms. New class of algorithms based on the functional performance model (FPM), representing the speed of the processor by a function of problem size, has been recently proposed. These algorithms cannot be however employed in self-adaptable applications because of very high cost of construction of the functional performance model. The paper presents a new class of parallel algorithms for highly heterogeneous HPC platforms. Like traditional FPM-based algorithms, these algorithms assume that the speed of the processors is characterized by speed functions rather than speed constants. Unlike the traditional algorithms, they do not assume the speed functions to be given. Instead, they estimate the speed functions of the processors for different problem sizes during their execution. These algorithms do not construct the full speed function for each processor but rather build and use their partial estimates sufficient for optimal distribution of computations with a given accuracy. The low execution cost of distribution of computations between heterogeneous processors in these algorithms make them suitable for employment in self-adaptable applications. Experiments with parallel matrix multiplication applications based on this approach are performed on local and global heterogeneous computational clusters. The results show that the execution time of optimal matrix distribution between processors is significantly less, by orders of magnitude, than the total execution time of the optimized application.

*Keywords:* parallel algorithms; heterogeneous platforms; data partitioning algorithms; functional performance models

## 1. Introduction

Conventional algorithms for distribution of computations between heterogeneous processors [1, 13] are based on a performance model which represents the speed of a processor by a constant positive number, and computations are distributed between the processors in proportion to this speed of the processor. The constant characterizing the performance of the processor is typically its relative speed demonstrated during the execution of a serial benchmark code solving locally the core computational task of some given size.

The fundamental assumption of the conventional algorithms based on the constant performance models (CPMs) is that the absolute speed of the processors does not depend on the size of the computational task. This assumption proved to be accurate enough if:

- The processors, between which we distribute computations, are all general-purpose ones of the traditional architecture,
- The same code is used for local computations on all processors, and
- The partitioning of the problem results in a set of computational tasks that are small enough to fit into the main memory of the assigned processors and large enough not to fit into the cache memory.

These conditions are typically satisfied when medium-sized scientific problems are solved on a heterogeneous network of workstations. Actually, heterogeneous networks of workstations were the target platform for the conventional heterogeneous parallel algorithms. However, this assumption of independence of the absolute speed of the processor on the size of the computational task becomes much less accurate in the following situations:

- The partitioning of the problem results in some tasks either not fitting into the main memory of the assigned processor and hence causing paging or fully fitting into faster levels of its memory hierarchy.
- Some processing units involved in computations are not traditional general-purpose processors (say, accelerators such as GPUs or specialized cores). In this case, the relative speed of a traditional processor and a non-traditional one may differ for two different sizes of the same computational task even if both sizes fully fit into the main memory.

---

[*] Corresponding author. Tel.: +353-1-716-2916; fax: +353-1-269-7262; e-mail: Alexey.Lastovetsky@ucd.ie.

- Different processors use different codes to solve the same computational problem locally.

The above situations become more and more common in modern and especially perspective high-performance heterogeneous platforms. As a result, applicability of the traditional CPM-based distribution algorithms becomes more restricted. Indeed, if we consider two really heterogeneous processing units $P_i$ and $P_j$, then the more different they are, the smaller will be the range $R_{ij}$ of sizes of the computational task where their relative speed can be accurately approximated by a constant. In the case of several different heterogeneous processing units, the range of sizes where CPM-based algorithms can be applied will be given by the intersection of these pair-wise ranges, $\bigcap_{i,j=1}^{p} R_{ij}$. Therefore, if a high-performance computing platform includes a relatively large number of significantly heterogeneous processing units, the area of applicability of CPM-based algorithms may become quite small or even empty. For such platforms, new algorithms are needed that would be able to optimally distribute computations between processing units for the full range of problem sizes.

The functional performance model (FPM) of heterogeneous processors proposed and analyzed in [16, 17] has proven to be more realistic than the constant performance models because it integrates many important features of heterogeneous processors such as the architectural and platform heterogeneity, the heterogeneity of memory structure, the effects of paging and so on. The algorithms employing it therefore distribute the computations across the heterogeneous processing units more accurately than the algorithms employing the constant performance models. Under this model, the speed of each processor is represented by a continuous function of the size of the problem. While this model is application centric because, generally speaking, different applications will characterize the speed of the processor by different functions, a speed function is supposed to satisfy some general restrictions on its shape [16]. In particular, beginning from some point, it should be monotonically decreasing.

The problem of distributing independent chunks of computations over a one-dimensional arrangement of heterogeneous processors using this functional performance model has been studied in [16]. It can be formulated as follows: Given $n$ independent chunks of computations, each of equal size (i.e., each requiring the same amount of work), how can we assign these chunks to $p$ ($p < n$) physical processors $P_1,\ldots,P_p$ with their respective full functional performance models represented by speed functions $s_1(x),\ldots,s_p(x)$ so that the workload is best balanced? An algorithm solving this problem with a complexity of $O(p \times \log_2 n)$ under the assumption of bounded heterogeneity of the processors is also proposed in [16]. This and other similar algorithms, which relax the restriction of bounded heterogeneity and are not sensitive to the shape of speed functions [15, 17], all are based on the following observation. The optimal data distribution points $(x_1, s_1(x_1)),\ldots,(x_p, s_p(x_p))$ lie on a straight line passing through the origin of the coordinate system and are the intersecting points of this line with the graphs of the speed functions of the processors: $\frac{x_1}{s_1(x_1)} = \frac{x_2}{s_2(x_2)} = \ldots = \frac{x_p}{s_p(x_p)}$, where $x_1 + x_2 + \ldots + x_p = n$. This is shown in Fig. 1 for $p = 4$. These algorithms are used as building blocks in algorithms solving more complicated linear algebra kernels such as the dense factorizations [14].

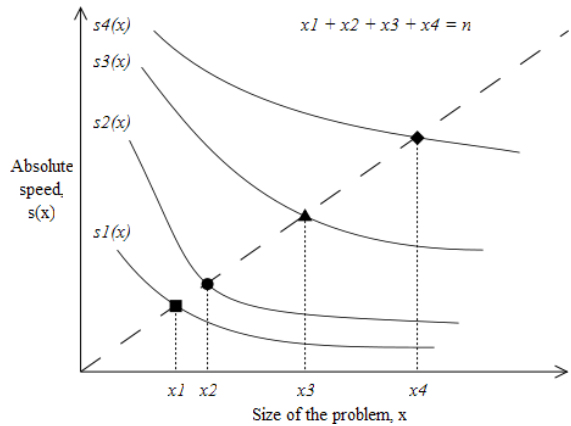

**Fig. 1. Optimal data distribution showing the geometric proportionality of the number of chunks to the speed of the processor.**

The cost of experimentally building the full functional performance model of a processor, i.e., the model for the full range of problem sizes, is very high. This is due to several reasons. To start with, the accuracy of the model depends on the number of experimental points used to build it. The larger the number, the more accurate the model is. However, there is a cost associated with obtaining an experimental data point, which requires execution of a computational kernel for a specified problem size. This cost is especially high for problem sizes in the region of paging. Also, the number of experimental points required to build the full functional performance model increases remarkably as the number of parameters used to represent the problem size increases, as shown in the experimental results in this paper.

The high model-construction cost limits the applicability of parallel algorithms based on full FPMs to situations where the construction of the full FPMs of heterogeneous processors and their use in the application can be separated. For example, if we develop an application for dedicated stable heterogeneous platforms, which is supposed to be executed on the same platform multiple times, we can build the full FPMs of the heterogeneous processors of the platform once and then use these models multiple times during the repeated execution of the application. In this case, the time of construction of the FPMs can become very small compared to the accumulated performance gains during the multiple executions of the optimized application. However, this approach does not apply to applications, each run of which is considered unique. This is the case for applications that are supposed to be executed in dynamic environments or any other environments where the number of available processors and their performance characteristics can be different for different runs of the same application. This is also the case for applications that can be run just once or a small number of times in each environment. Such applications should be able to optimally distribute computations between the processors of the executing platform assuming that this platform is different and *a priori* unknown for each run of the application. In this paper, we call applications that automatically adapt at runtime to any set of heterogeneous processors with a priori unknown performance characteristics *self-adaptable* applications.

The problem of minimization of the cost of construction of the full functional performance model of a processor has been studied recently proposing a relatively efficient sub-optimal solution [19]. However, even if an ideal optimal procedure becomes available to build approximations of the models of heterogeneous processors, the fact remains that the cost of construction of the full model is too high to forbid the use of data partitioning algorithms, employing the full model, in self-adaptable applications.

The paper addresses this problem and presents a new algorithm of distributing independent chunks of computations over a one-dimensional arrangement of heterogeneous processors suitable for the use in self-adaptable applications. Like the traditional FPM-based algorithms, the algorithm assumes that the speed of the processors is characterized by speed functions rather than speed constants. Unlike the traditional algorithms, it does not require the speed functions to be given. Instead, it estimates the speed functions of the processors for different problem sizes during its execution. This makes the algorithm distributed as its execution will involve all the heterogeneous processors. The algorithm does not construct the complete speed function for each processor but rather builds and uses its partial estimate sufficient for optimal distribution of computations. The proposed algorithm returns a solution not perfectly balancing the load of the processors but rather a solution balancing their load with a given accuracy.

Using experimental results for parallel matrix multiplication on local and global heterogeneous computational clusters, we demonstrate that the execution time of the proposed distributed mapping algorithm is orders of magnitude less than the total execution time of the optimized parallel application, thereby making it very suitable for employment in self-adaptable applications.

The rest of the paper is organized as follows. In Section 2, we present the main contribution of this paper, which is the distributed functional algorithm of mapping independent chunks of computations onto heterogeneous processors. This is followed by experimental results on local and global heterogeneous computing clusters in Section 3. We use a simple one-dimensional parallel matrix multiplication application for detailed cost analysis of the distributed functional algorithm. We use a two-dimensional matrix multiplication to show how to apply the method to realistic scientific applications and demonstrate the performance gains over traditional CPM-based applications. Related work is analysed in Section 4.

## 2. Distributed functional partitioning algorithm (DFPA)

In the mathematical form, the problem of distribution of computations over heterogeneous processors that we are trying to solve can be formulated as the following data partitioning problem:

**Given:**
- A set of $n$ independent units of computation each of equal size (i.e., each requiring the same amount of work);
- A set of $p$ ($p < n$) processors $P_1,\ldots,P_p$, whose speeds of processing $x$ units, $s_i = s_i(x)$, can be obtained by measuring the execution time, $t_i(x)$, of a computational kernel, $s_i(x) = x/t_i(x)$;
- A required relative accuracy of the solution, $\varepsilon$;

**To do:** partition the set of computation units into $p$ subsets so that:
- There is one-to-one mapping between the partitions and the processors, and

- $\max_{1 \leq i,j \leq p} \left| \frac{t_i(n_i) - t_j(n_j)}{t_i(n_i)} \right| \leq \varepsilon$, where $n_i$ is the number of computation units allocated to processor $P_i$ ($1 \leq i \leq p$).

Thus, the problem we study is to balance the load of heterogeneous processors with a given accuracy. The fundamental assumption, which makes efficient solution of this problem particularly difficult, is that the speeds of the processors are not known a priori. Therefore, if a partitioning algorithm needs the speed of processing of a given number of computation units by one or the other processor, it has to execute the corresponding number of units on this processor. Our solution to this problem is the following distributed data partitioning algorithm:

**Input:**
- $n$, the number of computation units;
- $p$ ($p < n$) processors $P_1,\ldots,P_p$;
- $\varepsilon$, the termination criterion.

**Output:**
- $d$, an integer array of size $p$, the $d_i$ is the number of computation units allocated to processor $P_i$;
- $t$, an integer array of size $p$, the $t_i$ is the execution time $t_i(d_i)$ observed on processor $P_i$ with $d_i$ computation units.

**Algorithm** (Fig. 2 illustrates the operation of the DFPA algorithm using an example with four heterogeneous processors $P_1, P_2, P_3, P_4$):

1. All the $p$ processors execute $n / p$ computation units in parallel. The execution times $t_1(n / p),\ldots,t_p(n / p)$ are gathered on processor $P_1$.

2. IF $\max_{1 \leq i,j \leq p} \left| \frac{t_i(n/p) - t_j(n/p)}{t_i(n/p)} \right| \leq \varepsilon$ THEN the even distribution of computations solves the problem and the algorithm stops;

   ELSE processor $P_1$ calculates the absolute speeds of the processors, $s_i(n / p) = (n / p) / t_i(n / p)$ for $1 \leq i \leq p$ and builds the first approximation of their functional performance models in the form of constant models, $s_i(x) = s_i(n / p)$, as illustrated in Fig. 2(a).

3. Using the data partitioning algorithm [16], processor $P_1$ calculates a new distribution of computation units, $d_1,\ldots,d_p$, which will be optimal for the current approximations of the functional performance models, and then sends a message to each processor $P_i$ informing the latter of its new allocation of computation units, $d_i$ (Fig. 2(a, c, e)).

4. Each processor $P_i$ then executes $d_i$ computation units in parallel with the other processors, $1 \leq i \leq p$. The execution times $t_1(d_1),\ldots,t_p(d_p)$ are gathered on processor $P_1$.

5. IF $\max_{1 \leq i,j \leq p} \left| \frac{t_i(d_i) - t_j(d_j)}{t_i(d_i)} \right| \leq \varepsilon$, THEN the current distribution of computation units, $d_1,\ldots,d_p$, solves the problem and EXIT;

   ELSE processor $P_1$ calculates the absolute speeds, which the processors demonstrated for this distribution of computation units, $s_i(d_i) = d_i / t_i(d_i)$ ($1 \leq i \leq p$), and uses these newly obtained points of the functional performance models of processors $P_i$, ($d_i, s_i(d_i)$), to build their more accurate piecewise linear approximations (as illustrated in Fig. 2(b, d, f)). Namely, let $\{(d_i^{(j)}, s_i(d_i^{(j)}))\}_{j=1}^{m}$, $d_i^{(1)} < \ldots < d_i^{(m)}$, be the experimentally obtained points of $s_i(x)$ used to build its current piecewise linear approximation, then
   - IF $d_i < d_i^{(1)}$, THEN the line segment $(0, s_i(d_i^{(1)})) \to (d_i^{(1)}, s_i(d_i^{(1)}))$ of this approximation will be replaced by two connected line segments $(0, s_i(d_{(i)})) \to (d_i, s_i(d_i))$ and $(d_i, s_i(d_i)) \to (d_i^{(1)}, s_i(d_i^{(1)}))$;
   - IF $d_i > d_i^{(m)}$, THEN the line $(d_i^{(m)}, s_i(d_i^{(m)})) \to (\infty, s_i(d_i^{(m)}))$ of this approximation will be replaced by the line segment $(d_i^{(m)}, s_i(d_i^{(m)})) \to (d_i, s_i(d_i))$ and the line $(d_i, s_i(d_i)) \to (\infty, s_i(d_i))$;
   - IF $d_i^{(k)} < d_i < d_i^{(k+1)}$, THEN the line segment $(d_i^{(k)}, s_i(d_i^{(k)})) \to (d_i^{(k+1)}, s_i(d_i^{(k+1)}))$ will be replaced by two connected line segments $(d_i^{(k)}, s_i(d_i^{(k)})) \to (d_i, s_i(d_i))$ and $(d_i, s_i(d_i)) \to (d_i^{(k+1)}, s_i(d_i^{(k+1)}))$.

6. GOTO 3.

In Fig. 2(a, c, e), the straight dashed line passing through the origin of the coordinate system is the result of the data partitioning algorithm [16] applied to the piecewise linear approximations of functional performance models, which are built at the previous step (the initial approximations are the constants $s_i(n / 4)$). Therefore, this line fluctuates from step to step, eventually coming close to the optimum (Fig. 2(e)). In Fig. 2(b, d, f), we also show how the line connecting the points ($d_i$, $s_i(d_i)$) evolves. In the beginning, this dotted dashed line is a curve. Gradually, it is straightening, which indicates that the load of processors is becoming more and more balanced (Fig. 2(f)).

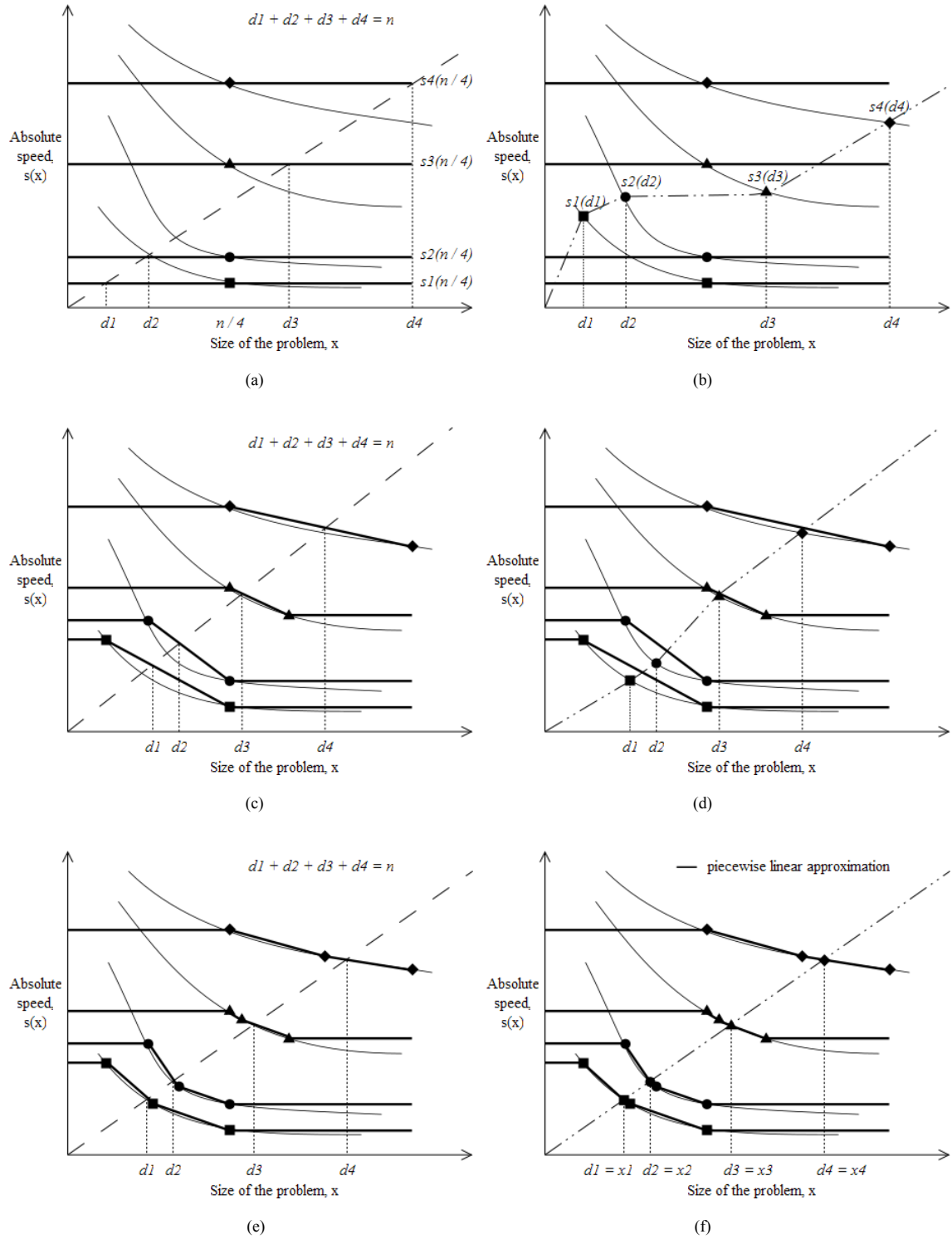

**Fig. 2.** Steps of the distributed functional partitioning algorithm (DFPA) illustrated using four heterogeneous processors.

In Fig. 2(a, c, e), the straight dashed line passing through the origin of the coordinate system is the result of the data partitioning algorithm [16] applied to the piecewise linear approximations of functional performance models, which are built at the previous step (the initial approximations are the constants $s_i(n / 4)$). Therefore, this line fluctuates from step to step, eventually coming close to the optimum (Fig. 2(e)). In Fig. 2(b, d, f)), we also show how the line connecting the points ($d_i$, $s_i(d_i)$) evolves. In the beginning, this dotted dashed line is a curve. Gradually, it is straightening, which indicates that the load of processors is becoming more and more balanced (Fig. 2(f)).

**Proposition.** *Given the full functional performance models of the processors $P_1,…,P_p$ satisfy the assumptions about their shape stated in [16], the DFPA algorithm always converges.*

Indeed, first of all, by construction, the piecewise linear approximations of the full functional performance models used in the algorithm will satisfy the same assumptions about their shape as the full models themselves. Therefore, at each iteration step, application of algorithm [16] to the set of approximate models will be successful and return the optimal solution for these approximate models. Second, each next iteration step of the algorithm results in more accurate approximation of the segments of the full models that contain the points of the optimal solution. Therefore, after a number of iterations, the approximations of the full functional performance models will become accurate enough in order algorithm [16] to return a solution sufficiently close to the optimal one.

## 3. Application and experimental results

This section consists of two sub-sections. The first one presents a detailed experimental cost analysis of DFPA. For this purpose, we use a very simple application, one-dimensional parallel matrix multiplication, which allows us to focus on different sources of overhead rather than on peculiarities of the implemented algorithm. In the second sub-section, we show how to apply the DFPA-based approach to design and implementation of a realistic scientific application, namely, the two-dimensional heterogeneous matrix multiplication, and demonstrate that this DFPA-based application outperforms its traditional CPM-based counterpart. The main goal of this sub-section is to demonstrate how DFPA can be used to improve performance of realistic applications on heterogeneous platforms rather than design of optimal matrix-multiplication algorithm (which is our ongoing research work).

All our experimental work is conducted on computational clusters consisting of traditional general-purpose heterogeneous processors. As we have discussed in Section 1, the partitioning algorithm proposed in this work is designed for the situations when execution of applications on the heterogeneous platform cannot be accurately described by constant performance models. In our experimental setup, this happens when the partitioning results in computational tasks that:

- Either fully fit in higher levels of the memory hierarchy (see Fig. 3 illustrating this case for 4 processors on one of our clusters), or
- Do not fit in the main memory and require paging.

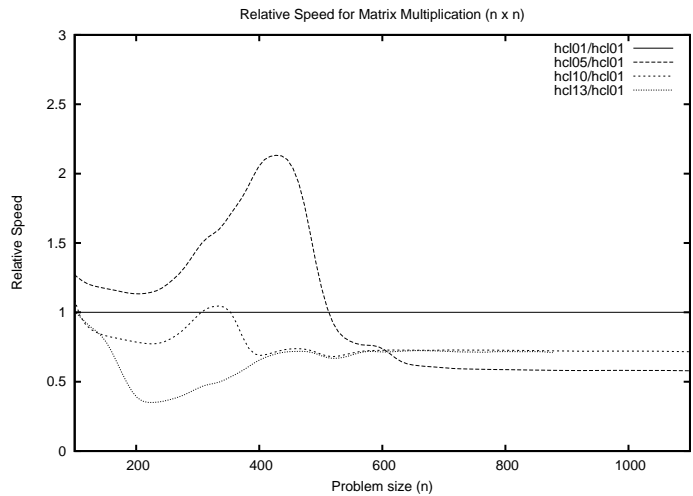

**Fig. 3. Relative processor speeds for a naive matrix multiplication in the ranges of cache and main memory.**

The paging region is the most expensive in terms of the cost of DFPA. Therefore, our cost analysis focuses on this region.

### 3.1. Cost analysis

We analyse the cost of DFPA with help of a heterogeneous application that implements matrix operation $C = A \times B$, multiplying matrix $A$ and matrix $B$, where $A$, $B$, and $C$ are dense square matrices of size $n \times n$, on a network of $p$ heterogeneous processors (see Fig. 4(a)). We use a 1D processor arrangement. The matrices $A$ and $C$ must be horizontally sliced so that the height of the slice is proportional to the speed of the processor owning the slice. All the processors contain all the elements of matrix $B$. We assume that only one process is configured to execute on a processor. We purposely choose an application with no communications because the goal of the experiments is not to show how to multiply matrices in parallel but to analyse the cost and demonstrate the practical speed of convergence of the distributed partitioning algorithm. In terms of relative performance of the DFPA, the results can only change in DFPA's favour for more complicated algorithms involving communications.

For this application, the core computational kernel performs a matrix update of a matrix $C_b$ of size $n_b \times n$ using $A_b$ of size $n_b \times 1$ and $B_b$ of size $1 \times n$ as shown in Fig. 4(b). The size of the problem is represented by two parameters, $n_b$ and $n$. The total number of matrix elements stored on each processor will be $(2 \times n_b \times n + n \times n)$. We use a combined computation unit, which is made up of one addition and one multiplication, to express the volume of computation. The total number of computation units needed to solve this problem will be equal to $n_b \times n$. Therefore, the absolute speed of the processor exposed by the application when solving the problem of size $(n_b, n)$ can be calculated as $n_b \times n$ divided by the execution time of the matrix update. This gives us a function, $f: \mathbf{N}^2 \to \mathbf{R}_+$, mapping problem sizes to speeds of the processor. The functional performance model of the processor is obtained by continuous extension of function $f: \mathbf{N}^2 \to \mathbf{R}_+$ to function $g: \mathbf{R}_+^2 \to \mathbf{R}_+$ ($f(n,m) = g(n,m)$ for any $(n,m)$ from $\mathbf{N}^2$).

For the first set of experiments, we used a 16-node heterogeneous Linux cluster with a Gigabit Ethernet switch and Open MPI. The cluster specifications are presented in Table 1. Fig. 5(a) depicts this function for one of the processors, **hcl11**, used in experiments. Fig. 5(b) shows the relative speed of two processors, **hcl09** and **hcl06**, calculated as the ratio of their absolute speeds. One can see that the relative speed varies significantly depending on the value of variables $x$ and $y$ (the variables represent $n_b$ and $n$). The heterogeneity of the processors is calculated as the ratio of the absolute speed of the fastest processor to the absolute speed of the slowest processor. For example, consider the benchmark code of the core computational kernel with $n_b = 20$ and $n = 2048$. The absolute speeds of the processors in million flop/s performing this update are {658, 667, 648, 644, 570, 503, 583, 581, 611, 628, 567, 601, 338, 651, 554, 695}. As one can see, **hcl16** is the fastest processor and **hcl13** is the slowest processor. The heterogeneity is therefore 2.

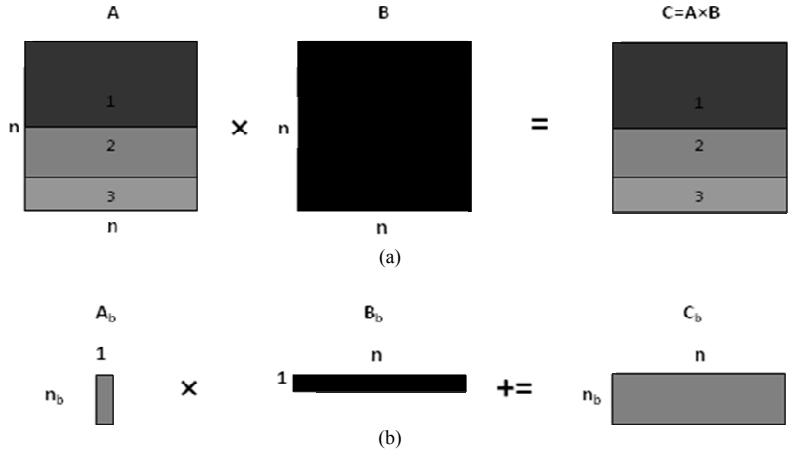

**Fig. 4.** (a) Matrix operation $C = A \times B$ on a network of three heterogeneous processors. Matrices $A$ and $C$ are horizontally sliced such that the height of the slice ($n_b$) is proportional to the speed of the processor. (b) The computational kernel for the second processor performs a matrix update of $A_b$ of size $n_b \times 1$ and $B_b$ of size $1 \times n$ to give a dense matrix $C_b$ of size $n_b \times n$.

**Table 1. Specifications of the HCL cluster.**

| Host | Model | Processor | Bus | L2 Cache | RAM |
|---|---|---|---|---|---|
| hcl01-hcl04 | Dell Poweredge 750 | 3.4 Xeon | 800MHz | 1MB | 1GB |
| hcl05 | Dell Poweredge SC1425 | 3.6 Xeon | 800MHz | 2MB | 256MB |
| hcl06 | Dell Poweredge SC1425 | 3.0 Xeon | 800MHz | 2MB | 256MB |
| hcl07-hcl08 | Dell Poweredge 750 | 3.4 Xeon | 800MHz | 1MB | 256MB |
| hcl09-hcl10 | IBM E-server 326 | 1.8 AMD Opteron | 1GHz | 1MB | 1GB |
| hcl11 | IBM X-Series 306 | 3.2 P4 | 800MHz | 1MB | 512MB |
| hcl12 | HP Proliant DL 320 G3 | 3.4 P4 | 800MHz | 1MB | 512MB |
| hcl13 | HP Proliant DL 320 G3 | 2.9 Celeron | 533MHz | 256KB | 1GB |
| hcl14 | HP Proliant DL 140 G2 | 3.4 Xeon | 800MHz | 1MB | 1GB |
| hcl15 | HP Proliant DL 140 G2 | 2.8 Xeon | 800MHz | 1MB | 1GB |
| hcl16 | HP Proliant DL 140 G2 | 3.6 Xeon | 800MHz | 2MB | 1GB |

We compare the efficiency of two matrix multiplication applications. The first application uses pre-built functional performance models of the processors to find the optimal partitioning of the matrices by executing the Full-Functional-Model Partitioning Algorithm (FFMPA) on a single processor. The found partitioning is then used to distribute the matrices between the processors and execute the parallel matrix multiplication itself. Like the first application, the second one finds the optimal distribution of the matrices over the processors and then uses this distribution for parallel matrix multiplication. But unlike the FFMPA-based application, it does not need the functional performance models of the processors. Instead, it executes the DFPA on all the processors. In all our experiments, the DFPA returned almost the same data distribution as the FFMPA.

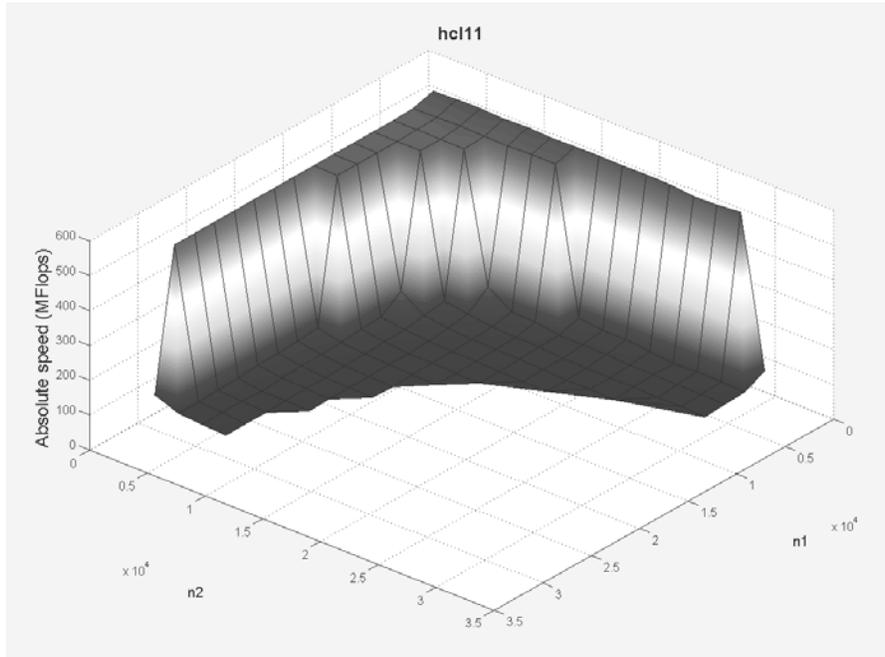

(a)

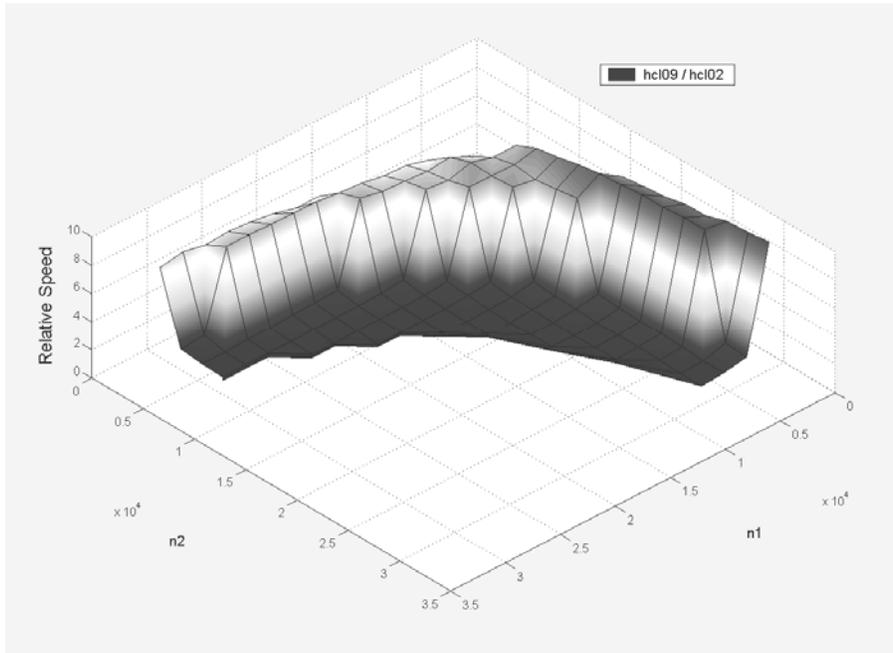

(b)

Fig. 5. (a) The absolute speed of a processor **hcl11** as a function of the size of the computational task of updating a dense x×y matrix.
(b) The relative speed of two processors (**hcl09**, **hcl06**) calculated as the ratio of their absolute speeds.

Table 2 shows the execution times of these two applications. The second column gives the execution time of the FFMPA-based application, which does not include the time of construction of the functional performance models of the processors. The third column shows the total execution time of the DFPA-based application, including the execution time of the DFPA. The DFPA execution time includes both its computation and communication costs. The ratio of the execution times of these two applications is presented in the fourth column.

**Table 2. Performance of the FFMPA and DFPA based applications on 15 processors of the HCL cluster (excl. hcl07).**

| Matrix size ($n \times n$) | FFMPA-based application, not including model building (sec) | DFPA-based application, including DFPA (sec) | DFPA-based / FFMPA-based | DFPA execution time (sec) | DFPA iterations |
|---|---|---|---|---|---|
| 2048 | 3.16 | 3.43 | 1.06 | 0.22 | 4 |
| 3072 | 10.70 | 11.02 | 1.02 | 0.30 | 2 |
| 4096 | 25.42 | 25.87 | 1.01 | 0.43 | 2 |
| 5120 | 52.61 | 57.62 | 1.09 | 4.96 | 11 |
| 6144 | 101.45 | 112.19 | 1.10 | 10.74 | 3 |
| 7168 | 183.79 | 203.36 | 1.10 | 19.55 | 5 |
| 8192 | 280.04 | 308.88 | 1.10 | 28.84 | 5 |

We want to stress again that the execution time of the FFMPA-based application does not include the time of building the full functional performance models of the processors. In our case, parallel building of the full models of the processors has taken 1850 seconds. The range of problem sizes, ($n_b$, $n$), used for building them, is $n_b = n / 80, 2n / 80, \ldots, n / 4$ and $n = 1024, 2048, \ldots, 8192$. One can see that this time is several orders of magnitude higher than the DFPA execution times shown in the fifth column. The number of experimental points used to build the full models for this range is $20 \times 8 = 160$. This is compared to a maximum of 11 points used by the DFPA (see column 6 of Table 1). Thus, we can conclude that the DFPA converges very fast and its contribution into the total execution time of the application has never exceeded 10%. It is also very efficient in terms of the number of experimental points.

Table 3 demonstrates the performance of the DFPA for two different values of the termination criterion $\varepsilon$, 10% and 2.5%. In both cases, we obtained very similar data distributions. The fourth and eight columns show that the number of DFPA iterations slightly increases with the decrease of the relative error, $\varepsilon$. However, the overall DFPA execution time does not increase significantly.

**Table 3. The DFPA-based application performance on 15 processors of the HCL cluster (excl. hcl07) with $\varepsilon$ = 10% and 2.5%.**

| Matrix size ($n \times n$) | 10% | | | 2.5% | | |
|---|---|---|---|---|---|---|
| | Matrix multiplication (sec) | DFPA (sec) | DFPA iterations | Matrix multiplication (sec) | DFPA (sec) | DFPA iterations |
| 2048 | 3.21 | 0.22 | 4 | 3.16 | 0.23 | 6 |
| 3072 | 10.72 | 0.30 | 2 | 10.70 | 0.31 | 3 |
| 4096 | 25.44 | 0.43 | 2 | 25.42 | 0.49 | 4 |
| 5120 | 52.66 | 4.96 | 11 | 52.61 | 6.18 | 11 |
| 6144 | 101.45 | 10.74 | 3 | 101.45 | 11.83 | 4 |
| 7168 | 183.81 | 19.55 | 5 | 183.79 | 21.05 | 5 |
| 8192 | 280.04 | 28.84 | 5 | 280.04 | 26.78 | 5 |

Let us consider the experiment with $n = 5120$ in more detail since in this case the DFPA takes more steps than usual, namely 11. In this case, several processors (including nodes **hcl06, hcl08**) operated in the borderline region between light and heavy paging. Fig. 6 demonstrates how the DFPA finds the optimal data distribution in this case. For the initial distribution ($n_b = 341$), the **hcl06, hcl08** nodes show very slow speed because of paging. On the rest of the processors, paging does not take place (see, for example, the speed of the **hcl03, hcl16** nodes in Fig. 6). Based on this observation, the DFPA will give smaller slices of the matrices ($n_b = 20$ and 31) to the slower processors while the other processors will receive larger parts of the matrices. At the next iteration, these slower processors do not experience the paging and execute the computational kernel significantly faster than the processors with bigger RAM. Therefore, the DFPA will redistribute the matrices in favour of the processors with smaller RAM. During a few first iterations, the DFPA finds the leftmost and rightmost points of the speed functions and then converges to the optimal distribution. In the experiments with $n > 5120$ (see Table 2), the number of iterations became smaller because the optimal distribution moved away from steep slopes of the speed functions.

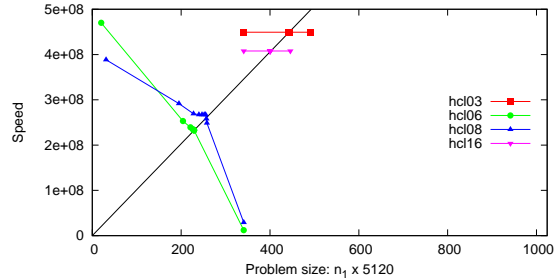

**Fig. 6. Execution steps of DFPA for $n = 5120$, $p = 15$, $\varepsilon = 2.5\%$ as observed on four most representative processors.**

We performed the same experiments on a larger heterogeneous platform, Grid5000 [10]. Grid5000 is a cluster of clusters geographically distributed between 8 sites in France. 14 types of nodes were involved in the experiments, with different number of processors and cores and different sizes of RAM. MPICH-1.2.7 and Gigabit Ethernet were used as a communication layer. The DFPA-based and FFMPA-based applications were run on 28 nodes, one process per node, for problem sizes $n = 7168, 10240, 12288$. The heterogeneity of processors was in the range of 2.5 to 2.8. The DFPA successfully and efficiently balanced the computational load (see Table 4). One can see that the DFPA took no more than 3 iterations and its contribution into the total execution time of the DFPA-based application did not exceed 1%.

**Table 4. The DFPA-based application performance on 28 nodes of Grid5000 with $\varepsilon = 10\%$ and 2.5%.**

| Matrix size ($n \times n$) | 10% | | | 2.5% | | |
|---|---|---|---|---|---|---|
| | Matrix multiplication (sec) | DFPA (sec) | DFPA iterations | Matrix multiplication (sec) | DFPA (sec) | DFPA iterations |
| 7168 | 65.88 | 1.19 | 2 | 65.71 | 1.24 | 3 |
| 10240 | 193.05 | 2.02 | 2 | 192.67 | 2.12 | 3 |
| 12288 | 334.32 | 2.65 | 2 | 333.87 | 2.74 | 3 |

### 3.2. Application of DFPA to matrix multiplication based on 2D heterogeneous partitioning

In this sub-section, we use DFPA to modify a heterogeneous two-dimensional matrix multiplication (Fig. 7(a)) based on the data partitioning scheme [13] (see Fig. 8), which in its turn is obtained by modification of the ScaLAPACK [3] two-dimensional matrix multiplication. The modification is that the heterogeneous two-dimensional data distribution of [13] is used instead of the standard homogeneous data distribution.

Further modification of this heterogeneous 2D matrix multiplication has been proposed in [18]. It employs the full functional performance model instead of the constant performance model for matrix partitioning. Unlike its CPM-based prototype, the FPM-based partitioning algorithm is iterative and can be summarized as follows. It starts with the matrix evenly partitioned into equal column panels. Then, at each iteration, (i) it uses the full FPM to optimally partition each column vertically. Then, (ii) the full FPM is used to calculate the sum of speeds of the processors along each column and re-partition the matrix in the horizontal dimension in proportion with these speeds. This way, this partitioning algorithm converges to the optimal partitioning. The full FPM is pre-built using the core computational kernel.

The core computational kernel for 2D matrix multiplication performs a matrix update of a matrix $C_b$ of size $m_b \times n_b$ using $A_b$ of size $m_b \times 1$ and $B_b$ of size $1 \times n_b$ as shown in Fig. 7(b). The size of the problem is represented by two parameters, $m_b$ and $n_b$. We use a combined computation unit, which is made up of one addition and one multiplication, to express the volume of computation. Therefore, the total number of computation units (namely, multiplications of two $b \times b$ matrices) performed during the execution of the benchmark code will be approximately equal to $m_b \times n_b$. Therefore, the absolute speed of the processor exposed by the application when solving the problem of size ($m_b,n_b$) can be calculated as $m_b \times n_b$ divided by the execution time of the matrix update. This gives us a function, f: $\mathbf{N}^2 \to \mathbf{R}_+$, mapping problem sizes to speeds of the processor. The FPM of the processor is obtained by continuous extension of function f: $\mathbf{N}^2 \to \mathbf{R}_+$ to function g: $\mathbf{R}_+^2 \to \mathbf{R}_+$ (f($n,m$) = g($n,m$) for any ($n,m$) from $\mathbf{N}^2$).

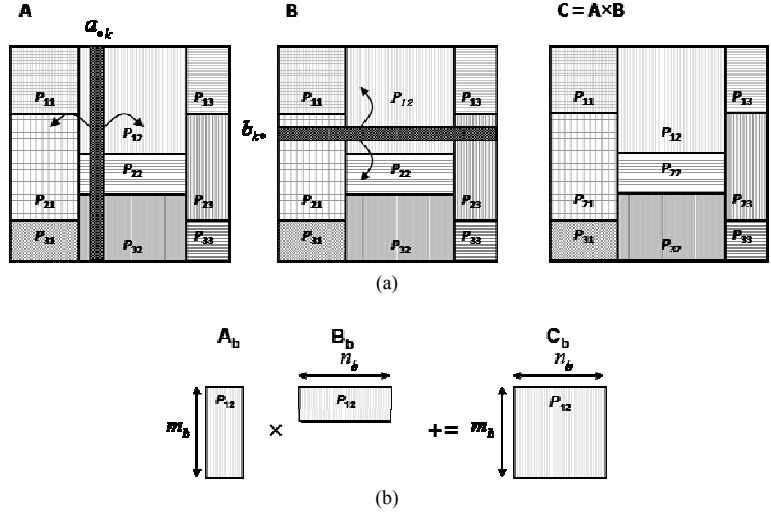

(a)

(b)

Fig. 7. (a) One step of the algorithm of parallel matrix multiplication employing a 2D heterogeneous processor grid of size 3×3. Matrices *A*, *B*, and *C* are partitioned so that the area of the rectangle is proportional to the speed of the processor owning it. First, each *b*×*b* block of the pivot column $a_{\bullet k}$ of matrix *A* (shown with curly arrows) is broadcast horizontally, and each *b*×*b* block of the pivot row $b_{k \bullet}$ of matrix *B* (shown with curly arrows) is broadcast vertically. Then, each *b*×*b* block $c_{ij}$ of matrix *C* is updated, $c_{ij}=c_{ij}+a_{ik}\times b_{kj}$. (b) The computational kernel (shown here for processor $P_{12}$ for example) performs a matrix update of a dense matrix $C_b$ of size $m_b \times n_b$ using $A_b$ of size $m_b \times 1$ and $B_b$ of size $1 \times n_b$. The matrix elements represent *b*×*b* matrix blocks.

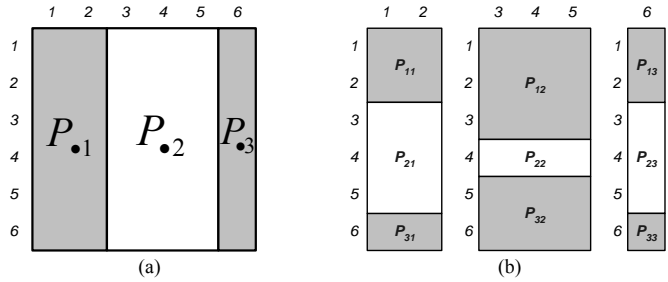

Fig. 8. Example of two-step distribution of a 6×6 square over a 3×3 processor grid. The relative speed of processors is given by {0.11, 0.25, 0.05, 0.17, 0.09, 0.08, 0.05, 0.17, 0.03}. (a) The 6×6 square is distributed in a one-dimensional block fashion over processors columns of the 3×3 processor grid in proportion 0.33:0.51:0.16 ≈ 2:3:1. (b) Each vertical rectangle is distributed independently in a one-dimensional block fashion over processors of its column. The first rectangle is distributed in proportion 0.11:0.17:0.05 ≈ 2:3:1. The second one is distributed in proportion 0.25:0.09:0.17 ≈ 3:1:2. The third is distributed in proportion 0.05:0.08:0.03 ≈ 2:3:1.

Construction of the full 2D FPM for this kernel is very expensive. The cost of one experimental point may be especially high for the problem sizes in the region of paging. We did propose DFPA and dynamically built partial FPMs as an efficient solution of this problem. Now we present a DFPA-based modification of the partitioning algorithm [18]. In this modification, DFPA is used for estimation of 1D projections of the full FPM. The projection is obtained by fixing the width of the column the vertical partitioning of which is performed (see Fig. 9). The partitioning of the columns (step (i) of the algorithm [18]) will now include partial estimation of the FPM. The obtained estimates are used in the horizontal re-partitioning (step (ii) of the algorithm [18]).

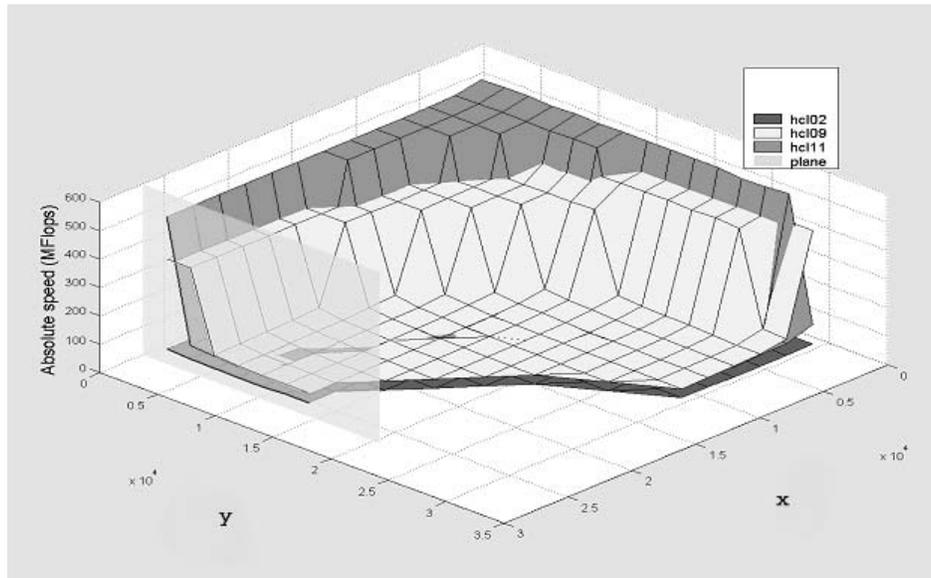

(a)

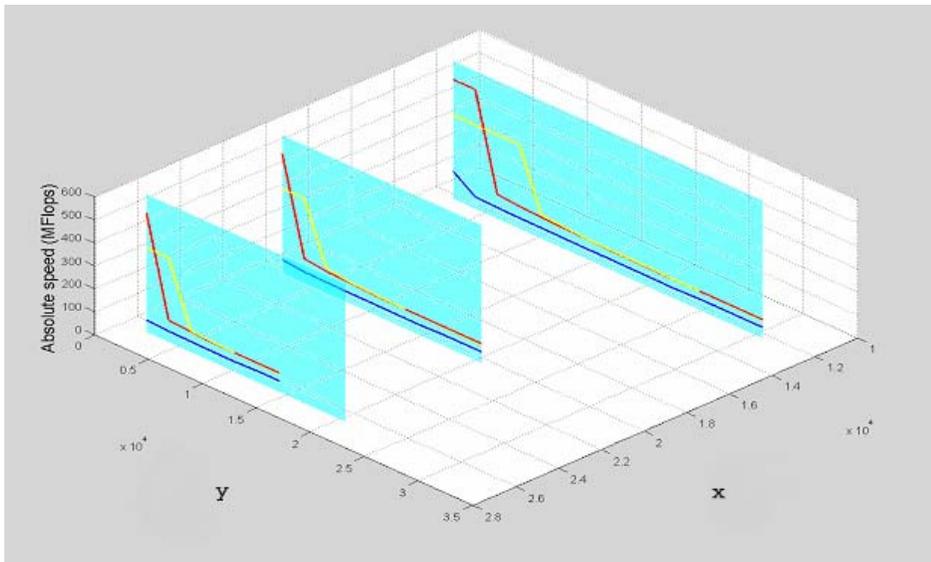

(b)

**Fig. 9. (a) Absolute speeds of 3 processors, $g_i(x,y)$. (b) Projections of absolute speeds of the processors, with column width fixed $x =$ 1.22, 2.02, 2.64 × $10^4$.**

Thus, the DFPA-based 2D matrix partitioning algorithm will be a nested iterative procedure:
- 2D partitioning that includes steps (i), (ii) (outer loop), and
- DFPA in step (i) (inner loop).

Formally, it can be summarized as follows:
1. Initially, matrices are partitioned evenly, with column widths $n_j = n / q$ and row heights $m_{ij} = m / p$, where $1 \leq i \leq p$, $1 \leq j \leq q$.

2. For each column $j$, the iterative DFPA procedure is executed in parallel, in order to estimate the FPM and optimally partition the rows in the column (step (i)). The results of this operation, the optimal row distributions, $(m_{1j},...,m_{pj})$, and the observed execution times, $(t(m_{1j},n_j),..., t(m_{pj},n_j))$, are gathered at the root processor, $P_{11}$, which executes step 3.

3. IF $\max\limits_{\substack{1 \leq i, x \leq p \\ 1 \leq j, y \leq q}} \left| \dfrac{t_{ij}(m_{ij},n_j) - t_{xy}(m_{xy},n_y)}{t_{ij}(m_{ij},n_j)} \right| \leq \varepsilon$, THEN the current distribution of computation units, $(m_{ij},n_j)$, where $1 \leq i \leq p$, $1 \leq j \leq q$, solves the problem and EXIT;

ELSE processor $P_{11}$ calculates the absolute speeds, which the processors demonstrated for this distribution of computation units, $s_{ij}(m_{ij},n_j) = m_{ij} \times n_j / t_{ij}(m_{ij},n_j)$, where $1 \leq i \leq P$, $1 \leq j \leq q$, and uses them to balance column widths (step (ii)) as follows: $n_j = n \dfrac{\sum\limits_{1 \leq i \leq p} s_{ij}(m_{ij},n_j)}{\sum\limits_{1 \leq y \leq q}\sum\limits_{1 \leq i \leq p} s_{iy}(m_{iy},n_y)}$. In other words, the new column width is proportional to the sum of relative speeds of the processors belonging to this column.

4. GOTO 2

Now we compare the efficiency of three 2D matrix multiplication applications, which use different partitioning algorithms. The first application executes partitioning algorithm based on CPMs, which are built from single benchmarks for each column width. The second application executes the Full-Functional-Model Partitioning Algorithm (FFMPA) [18] based on full functional performance models of the processors. The third, DFPA-based, application executes the partitioning algorithm described above. For local computations, all these applications use out-of-core algorithms, namely, the high performance GotoBLAS2 library [9].

Since the FFMPA-based application uses a pre-built FPM and does not perform any benchmarks, it demonstrates the best performance (Fig. 10). Unlike the FFMPA-based application, the DFPA-based one executes benchmarks during partial estimation of the FPM. The smaller the number of these benchmarks, the closer the performance of the DFPA-based application to the performance of the FFMPA-based one. Both partitioning algorithms based on FPM return similar matrix distributions, while the CPM-based algorithm is less accurate and results in slower matrix multiplication. Fig. 10 demonstrates that the CPM-based application with fewer benchmarks but non-optimal matrix distribution will be 25% slower than the DFPA-based application, which uses larger number of executions of the core computational kernel to build the performance model but results in much faster matrix multiplication itself due to more optimal matrix distribution.

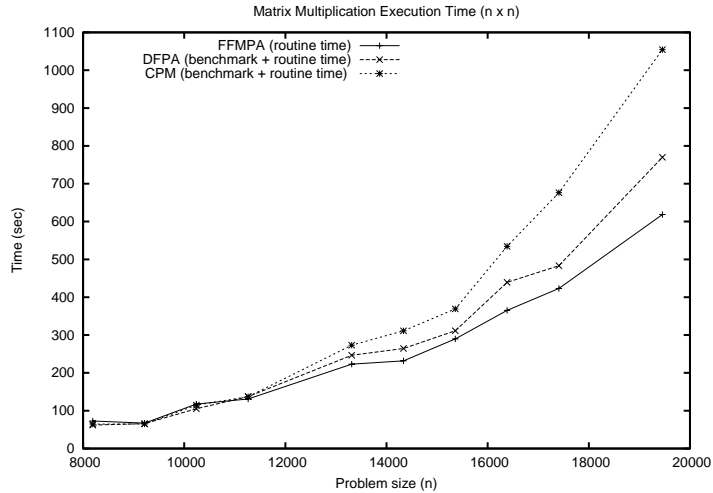

**Fig. 10. Performance of three heterogeneous 2D matrix multiplication applications on 16 nodes of the HCL cluster: CPM, based on the constant performance models; FFMPA, based in the pre-built full functional models; and DFPA, based on the dynamically built partial functional models.**

In implementation of the DFPA-based application, we used some optimization techniques in order to reduce both the number of benchmarks and the execution time of the most expensive ones. First of all, we use the results of all previous benchmarks when constructing the current estimate of the full FPM. Secondly, we do not change the width of the column if it is close enough to the previous width. Then, at the step 2 of the 2D partitioning algorithm, when DFPA is executed, we use not the equal row heights as the initial partitioning for DFPA but the row heights from the previous iteration of the 2D partitioning algorithm. This allows us to start benchmarking in the neighborhood of the 2D data distribution obtained at the previous iteration and, therefore, to reduce the probability of paging. Finally, we apply some low-level techniques to terminate some long-running benchmarks as soon as we get enough information to plot their results on the FPM estimate. The cost of DFPA-based 2D matrix partitioning on the HCL cluster is presented in Table 5.

Table 5. Performance of the DFPA-based heterogeneous 2D matrix multiplication application on the HCL cluster (16 nodes).

| Matrix size ($n \times n$) | Total execution time (sec) | DFPA time (sec) | DFPA iterations | Matrix multiplication (sec) | DFPA cost (%) |
|---|---|---|---|---|---|
| 8192 | 61.91 | 0.17 | 16 | 61.74 | 0.28 |
| 9216 | 65.91 | 0.14 | 11 | 65.76 | 0.21 |
| 10240 | 105.22 | 0.19 | 13 | 105.02 | 0.18 |
| 11264 | 137.34 | 0.22 | 15 | 137.11 | 0.16 |
| 13312 | 246.49 | 5.84 | 44 | 240.65 | 2.36 |
| 14336 | 264.45 | 16.25 | 62 | 248.20 | 6.14 |
| 15360 | 311.28 | 24.06 | 69 | 287.22 | 7.73 |
| 16384 | 448.27 | 28.44 | 71 | 419.83 | 6.34 |
| 17408 | 483.23 | 52.51 | 69 | 430.71 | 10.86 |
| 19456 | 770.00 | 131.45 | 74 | 638.55 | 17.07 |

## 4. Related work

A number of algorithms of parallel solution of scientific and engineering problems on heterogeneous processor have been designed and implemented [1, 5, 6, 13]. They use different performance models but all the models represent the speed of a processor by a single positive number, and computations are distributed over the processors so that their volume is proportional to this speed of the processor. Cierniak, Zaki and Li [4] use the notion of normalized processor speed (NPS) in their machine model to solve the problem of scheduling parallel loops at compile time for heterogeneous processors. NPS is a single number and is defined as the ratio of time taken to execute on the processor under consideration, with respect to the time taken on a base processor. In [1] and [22], normalized cycle-times are used, i.e. application dependent elemental computation times, which are computed via small-scale experiments (repeated several times, with an averaging of the results).

While the general functional performance model used in this paper is universally applicable to heterogeneous processing units, there are some situations when its more restricted versions can be used. For example, some carefully designed scientific codes employ out-of-core algorithms to efficiently use memory hierarchy. They demonstrate quite a sharp and distinctive curve of dependence of the absolute speed on the task size. Their design manages to delay the appearance of page faults, minimizing their number for medium task sizes. However, because any design can only delay paging but not avoid it, the number of page faults will start growing like a snowball beginning from some threshold value of the task size. For such codes, the speed of the processor can be approximated by a unit step function accurately enough. Such a model was studied in [7]. More precisely, under this model the performance of the processor is characterized by the execution time of the task, represented by a piecewise linear function of its size. The problem of *asymptotically optimal* partitioning a set was formulated with this model as a linear programming problem. As such, it relies on the state of the art in linear programming. Currently, little is known (especially, theoretically) about practical and efficient algorithms for the linear programming problem. Moreover, if the unknown variables in the linear programming problem are all required to be integers, then the problem becomes an integer linear programming problem, which is known to be NP-hard in many practical situations. This makes the use of the linear programming model in design of self-adaptable applications a very challenging problem that is still wide open. At the same time, in our experiments, we did employ scientific codes using out-of-core algorithms for local computations. The results have shown the general functional performance model and algorithms and applications based on it are very efficient in this case as well outperforming the traditional CPM-based algorithms and applications.

The algorithm proposed in this work is iterative using the speeds of heterogeneous processors observed at previous steps for redistribution of the workload for the next step and gradually finding the optimal workload distribution. This approach has been used extensively in scheduling and load balancing on heterogeneous platforms [8, 12, 20, 21]. Traditionally, the performance of processors is assumed constant when redistributing the workload. One advanced load balancing strategy, the task queue model implemented in [2], has used adaptive speed measurements, which allowed it to outperform the traditional model [11] based on single speed measurements.

## 5. Conclusion and future work

The current trend in high performance computing platforms is that they employ increasingly heterogeneous processing units, the relative speed of which cannot be accurately described by constant performance models, CPMs. Therefore, CPM-based parallel algorithms become less and less applicable on these platforms. The functional performance model, FPM, and FPM-based partitioning algorithms were proposed to address this issue. They proved to be able to efficiently solve scientific problems on highly heterogeneous platforms given the full functional performance model of the processing units is constructed and provided as an input parameter. These algorithms however cannot be directly used in self-adaptable applications due to a very high cost of construction of full FPMs. In this paper, we have proposed a new highly efficient FPM-based data partitioning algorithm that does not require the FPM as its input parameter. Instead, it builds a partial estimate of the FPM sufficient for optimal data partitioning with a given accuracy. We have experimentally demonstrated the low cost of this algorithm that allows its employment in self-adaptable applications. We have also shown how to apply this algorithm to implementation of realistic scientific applications that significantly outperform their traditional CPM-based counterparts.

Our ongoing research concerns the design and implementation of a self-adaptable highly-efficient parallel matrix multiplication code for a wide range of modern and perspective high-performance heterogeneous platforms.

## Acknowledgements

This publication has emanated from research conducted with the financial support of Science Foundation Ireland under Grant Number 08/IN.1/I2054.## References

[1] O. Beaumont, V. Boudet, F. Rastello, Y. Robert, Matrix Multiplication on Heterogeneous Platforms, IEEE Trans. Parallel Distrib. Syst. 12 (2001) 1033-1051.
[2] R. Cariño, I. Banicescu, Dynamic Load Balancing with Adaptive Factoring Methods in Scientific Applications, J. Supercomput. 44 (2008) 41-63.
[3] J. Choi, J. Demmel, I. Dhillon, J. Dongarra, S. Ostrouchov, A. Petitet, K. Stanley, D. Walker, R.C. Whaley, ScaLAPACK: A Portable Linear Algebra Library for Distributed Memory Computers – Design Issues and Performance, Comput. Phys. Commun. 97 (1996) 1-15.
[4] M. Cierniak, M. Zaki, W. Li, Compile-Time Scheduling Algorithms for Heterogeneous Network of Workstations, Computer J. 40 (1997) 356-372.
[5] P. Crandall, M. Quinn, Problem Decomposition for Non-Uniformity and Processor Heterogeneity, J. Braz. Comp. Soc. 2 (1995) 13-23
[6] E. Dovolnov, A. Kalinov, S. Klimov, Natural Block Data Decomposition for Heterogeneous Clusters, in: IPDPS 2003, IEEE Computer Society, 2003, p. 102a.
[7] M. Drozdowski, P. Wolniewicz, Out-of-Core Divisible Load Processing, IEEE Trans. Parallel Distrib. Syst. 14 (2003) 1048-1056.
[8] I. Galindo, F. Almeida, J. Badía-Contelles, Dynamic Load Balancing on Dedicated Heterogeneous Systems, in: EuroPVM/MPI 2008, LNCS 5205, Springer, 2008, pp. 64-74.
[9] K. Goto, R.A. van de Geijn, Anatomy of High-Performance Matrix Multiplication, ACM Trans. Math. Soft. 34 (2008) 1-25.
[10] http://www.grid5000.fr
[11] S. Hummel, J. Schmidt, R. Uma, J. Wein, Load-sharing in Heterogeneous Systems via Weighted Factoring, in: SPAA'96, ACM, 1996, pp. 318-328.
[12] S. Ichikawa, S. Yamashita, Static Load Balancing of Parallel PDE Solver for Distributed Computing Environment, in: PDCS-2000, ISCA, 2000, pp. 399-405.
[13] A. Kalinov, A. Lastovetsky, Heterogeneous Distribution of Computations Solving Linear Algebra Problems on Networks of Heterogeneous Computers, J. Parallel Distrib. Comput. 61 (2001) 520-535.
[14] A. Lastovetsky, R. Reddy, Data distribution for dense factorization on computers with memory heterogeneity, Parallel Comput. 33 (2007) 757-779.
[15] A. Lastovetsky, R. Reddy, Data Partitioning for Multiprocessors with Memory Heterogeneity and Memory Constraints, Sci. Program. 13 (2005) 93-112.
[16] A. Lastovetsky, R. Reddy, Data Partitioning with a Functional Performance Model of Heterogeneous Processors, Int. J. High Perform. Comput. Appl. 21 (2007) 76-90.
[17] A. Lastovetsky, R. Reddy, Data Partitioning with a Realistic Performance Model of Networks of Heterogeneous Computers, in: IPDPS'04, vol. 2, IEEE Computer Society, 2004, p. 104b.

[18] A. Lastovetsky, R. Reddy, Two-dimensional Matrix Partitioning for Parallel Computing on Heterogeneous Processors Based on their Functional Performance Models, in: HeteroPar 2009, LNCS 6043, Springer, 2010, pp. 112-121.
[19] A. Lastovetsky, R. Reddy, R. Higgins, Building the Functional Performance Model of a Processor, in: ACM SAC 2006, ACM Press, 2006, pp.746-753.
[20] A. Legrand, H. Renard, Y. Robert, F. Vivien, Mapping and Load-balancing Iterative Computations, IEEE T. Parall. Distr. 15 (2004) 546-558.
[21] J. Martínez, E. Garzón, A. Plaza, I. García, Automatic Tuning of Iterative Computation on Heterogeneous Multiprocessors with ADITHE, J. Supercomput. (published online November 05, 2009).
[22] A. Petitet, J. Dongarra, Algorithmic Redistribution Methods for Block-Cyclic Decompositions, IEEE Trans. Parallel Distrib. Syst.10 (1999)1201-1216.